\documentclass{optica-article}

\journal{opticajournal}

\articletype{Research Article}

\usepackage{lineno}
\usepackage{tabularx}
\usepackage{xcolor}
\usepackage{soul}
\usepackage{booktabs}  
\usepackage{caption}  
\usepackage{hyperref} 
\usepackage{xurl} 
\usepackage{jabbrv}
\usepackage[ruled,vlined]{algorithm2e}
\usepackage{amsfonts}
\newcommand{\RR}{\mathbb{R}}
\newcommand{\TT}{\mathcal{T}}
\newcommand{\PP}{\mathcal{P}}
\newcommand{\FF}{\mathcal{F}}
\newcommand{\GG}{\mathcal{G}}
\setstcolor{green}
\setulcolor{red}

\begin{document}


\title{Fast Non-Line-of-Sight Transient Data Simulation and an Open Benchmark Dataset}

\author{Yingjie Shi,\authormark{1,2,4} Jinye Miao,\authormark{1,2} Taotao Qin, \authormark{3} Fuyao Cai,\authormark{1,2}
Yi Wei,\authormark{1,2}Lingfeng Liu,\authormark{1,2}Tongyao Li,\authormark{1,2}Chenyang Wu,\authormark{1,2}Huan Liang,\authormark{1,2}
Yuyang Yin,\authormark{1,2}Lianfa Bai,\authormark{1,2}Enlai Guo,\authormark{1,2,5} and Jing Han,\authormark{1,2,6}
}

\address{\authormark{1}State key Laboratory of Extreme Environment Optoelectronic Dynamic Testing Technology and Instrument,Nanjing University of Science and Technology, NanJing210094\\
\authormark{2}Jiangsu Key Laboratory of Visual Sensing \& Intelligent·Perception, Nanjing University of Science and Technology, Nanjing, Jiangsu 210094, China\\
\authormark{3}National key laboratory of transient physics, Nanjing University of Science and Technology, Nanjing, Jiangsu 210094, China
}

\email{\authormark{4}syj@njust.edu.cn\\
\authormark{5}njustgel@njust.edu.cn\\
\authormark{6}eohj@njust.edu.cn} 


\begin{abstract}
Non-Line-of-Sight (NLOS) imaging reconstructs the shape and depth of hidden objects from picosecond-resolved transient signals, offering potential applications in autonomous driving, security, and medical diagnostics. However, current NLOS experiments rely on expensive hardware and complex system alignment, limiting their scalability. This manuscript presents a simplified simulation method that generates NLOS transient data by modeling light-intensity transport rather than performing conventional path tracing, significantly enhancing computational efficiency. All scene elements, including the relay surface, hidden target, stand-off distance, detector time resolution, and acquisition window are fully parameterized, allowing for rapid configuration of test scenarios. Reconstructions based on the simulated data accurately recover hidden geometries, validating the effectiveness of the approach. The proposed tool reduces the entry barrier for NLOS research and supports the optimization of system design.
\end{abstract}

\section{Introduction}

In practical optical sensing, occlusions are nearly unavoidable. Whether it is a pedestrian around a street corner or an internal component of complex machinery, once the target exits the direct line of sight, conventional imaging methods, which based on direct or single-bounce reflections, can scarcely provide geometric or radiometric information. Non-Line-of-Sight (NLOS) imaging offers a fundamental breakthrough. A laser pulse or natural light first illuminates a visible relay surface, such as a rough wall or scattering screen, whose diffuse reflections redirect photons into the hidden area. Light reflected from the occluded object then returns to the relay surface and subsequently to the detector, thereby enabling the reconstruction of objects that are not directly visible
\cite{faccio2020non,o2018confocal,heide2019non,cao2022high,shi2022non,shi2023imaging,miao2023super}. 

Existing approaches can be broadly categorized into two types. The first type spatially or phase-modulates the incident wavefront, or leverages statistical correlations in the speckle pattern, to infer object contours from a limited number of two-dimensional images
\cite{metzler2020deep,liu2023single,seidel2019corner,baradad2018inferring,yedidia2019using}. 
Due to the absence of absolute time-of-flight information, these methods typically generate angular projections on the visible wall but lack accurate depth recovery. The second type utilizes picosecond laser pulses in combination with single-photon avalanche diodes (SPADs) and time-correlated single-photon counters (TCSPCs) to record histograms of photon arrival times after a triple reflection involving the wall and the hidden object. Algorithms based on this data can reconstruct both the three-dimensional shape and achieve depth resolutions on the order of centimetres or millimetres
\cite{kirmani2011looking,kadambi2016occluded,pandharkar2011estimating,xu2018revealing,liu2019non}. 
However, this approach requires expensive equipment, including ultrafast lasers, low-noise high-frame-rate SPADs, precision timing electronics, and complex calibration procedures. As a result, experimental platforms capable of producing high signal-to-noise-ratio triple-bounce data are rare. Consequently, most new NLOS reconstruction algorithms can only be validated on a small set of public datasets or limited self-collected samples, restricting comprehensive performance evaluation across diverse scenes and noise conditions and hindering the broader adoption of NLOS technology
\cite{liu2021non,shi2023steady,zhang2024real}.

With the advancement of GPU parallelism and the graphics community’s increasing understanding of transient rendering, an increasing number of studies now simulate the time-resolved signals that a NLOS system would detect entirely in software
\cite{owens2008gpu,yi2021differentiable,plack2023fast,royo2022non,iseringhausen2020non,pediredla2024time}. 
In this context, “transient” differs from conventional radiance rendering. At picosecond temporal resolution, each time bin corresponds to only a few millimeters of optical path length, requiring every ray interaction to be precisely mapped onto the temporal axis. Consequently, the renderer must not only record each photon's scattering direction, energy, and wavelength, but also accumulate its total path length and assign each trajectory to the appropriate time-of-flight histogram.

Several offline transient renderers have been released as open-source tools, most of which adopt Monte Carlo frameworks based on ray tracing or bidirectional path tracing, and accelerate convergence through techniques such as photon beams, temporal filtering, and importance sampling. For NLOS scenarios, supplementary modules are often used to extend existing renderers with invisible relay surfaces and multi-bounce configurations. Nevertheless, these pipelines remain fully dependent on path tracing, making image quality highly sensitive to sample count and source distribution. More critically, any modification to the wall–target–detector geometry frequently requires substantial alterations to scene files or even rewriting of shaders and integrators, significantly limiting flexibility in system optimization and algorithm testing\cite{chen2020learned,chen2019steady}. Furthermore, such approaches typically require high-performance hardware to maintain acceptable rendering times, necessitate scene reconstruction for layout changes, and provide limited support for simulating detector noise and temporal jitter.

A computationally efficient and conceptually transparent simulation framework for NLOS sampling is presented, which is openly available for both academic and industrial applications. The simulator requires only a single depth map of the hidden object and provides a concise yet versatile parameter interface. Users can independently configure the lateral dimensions and spatial sampling density of the relay surface, its stand-off distance from the hidden target, and the temporal resolution and acquisition window of the detector. By adjusting these parameters, physically consistent transient measurements can be rapidly generated across scales ranging from meters to centimeters, without the need to write shaders or modify complex scene files. The extensibility of the framework enables rapid prototyping, system-level sensitivity analysis, and the development of standardized benchmarks for evaluating diverse reconstruction algorithms. This substantially lowers the entry barrier to NLOS research and facilitates the co-development of hardware and computational inversion techniques. Furthermore, transient datasets for seven object categories from the ShapeNet dataset have been simulated, and the reconstruction performance of several existing algorithms has been systematically evaluated using this dataset.

\section{Theory}

\begin{figure}[ht!]
    \centering\includegraphics[width=12cm]{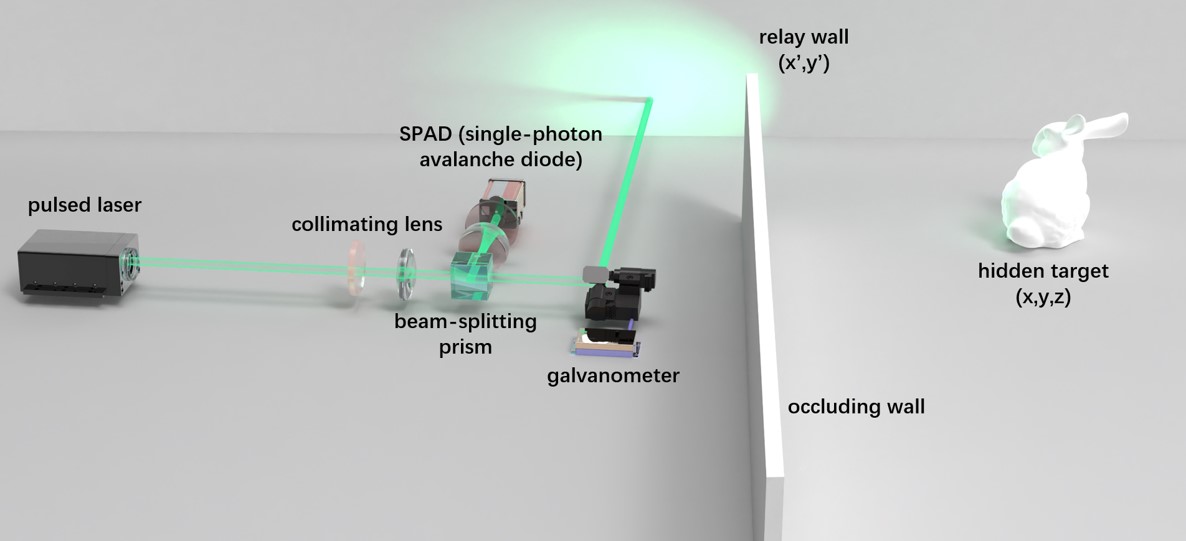}
    \caption{\textbf{Schematic diagram of the non‑line‑of‑sight imaging optical system}}
    \label{fig.1}
\end{figure}

In NLOS imaging, a confocal configuration, in which the illumination spot and the detection point coincide on the relay wall, offers high compatibility and scalability. Fig. \ref{fig.1} illustrates the optical layout used throughout this work. A pulsed laser is expanded by a collimating lens, pass through a beam‑splitting prism, and steered by a two‑axis galvanometer scanner to successive locations $(x',y')$ on a planar relay wall. Photons returning from the same point are routed by the prism onto a SPAD , thereby realising a confocal scan. An occluding wall blocks the direct line of sight to the hidden target $(x,y,z)$, so that only multiply scattered light reaches the detector.  

Within this setting, the light-cone 
transform (LCT) achieves an effective balance between physical accuracy 
and computational efficiency \cite{o2018confocal}, and is therefore adopted as the core model 
in our simulator. For a detection point $(x',y')$ on the relay wall, whose axial 
coordinate is set to $(z'=0)$ , the three-dimensional transient measurement
$\tau \in \mathbb{R}_{+}^{\left(x^{\prime}  \times  y^{\prime} \times t\right)}$
recorded from a Lambertian hidden object can be written as:
\[
  \tau(x',y',t)
  = \iiint_{\Omega} 
    \frac{1}{r^{4}}\,
    \rho(x,y,z)\,
    \delta \Bigl(
      2\sqrt{(x'-x)^{2}+(y'-y)^{2}+z^{2}} - t\,c
    \Bigr)
    \;dx\,dy\,dz,
\]
where $\rho \in \mathbb{R}_{+}^{\left(x  \times  y \times t\right)}$
denotes the albedo of the hidden surface at spatial coordinates $(x,y,z)$, $c$ 
is the speed of light, and $\tau(x',y',t)$ is the photon count recorded at wall position
$(x',y')$ at time $t$, the Dirac delta term describes a four-dimensional space-time hypercone 
that models the round-trip path of light from the relay wall to the target surface and 
back. By suitable simplification, this expression can be recast in a more compact form:
\[
  \tau(x',y',t)
  = \iiint_{\Omega}
      \frac{ \rho(x,y,z)\:
              \delta \Bigl(
                (x'-x)^{2} + (y'-y)^{2} + z^{2}
                - \bigl(t\,c/2\bigr)^{2}
              \Bigr)}
           {r^{3}}
      \;dx\,dy\,dz,
\]
setting $z = \sqrt{u}$, and $v = \left(\dfrac{t\,c}{2}\right)^{2}$ , the expression becomes
\[
  v^{\frac{3}{2}}\,
  \tau \left(x',y',\frac{c}{2}\sqrt{v}\right)
  \;=\;
  \iiint_{\Omega}
    \frac{ \rho \bigl(x,y,\sqrt{u}\bigr)\,
           \delta \Bigl( (x'-x)^{2} + (y'-y)^{2} + u - v \Bigr) }
         { 2\sqrt{u} }
    \;dx\,dy\,du.
\]
It can further be reduced to a three-dimensional convolution model:
\[
  \tau
  \;=\;
  R_{t}^{-1} \left\{
      R_{z}(p)\,* \operatorname{PSF}
  \right\},
\]
where $*$ denotes convolution,$R_{z}\{ \}$ resamples $p$ along the z-axis 
and multiplies the result by $\dfrac{1}{2\sqrt{u}}$, $R_{t}^{-1}\{ \}$ resamples 
$\tau$ along the temporal axis and scales it by $v^{3/2}$, and the point-spread function 
$PSF \in \mathbb{R}_{+}^{\left(x  \times  y \times t\right)}$ serves as the three-dimensional convolution kernel.

For computational convenience, the necessary notation and scaling are introduced. Let the target region in the three-dimensional positive real space be denoted as
$\Omega \in \mathbb{R}_{+}^{\,N\times N\times M}$, 
where, $N$ represents the physical extent, or equivalently the number of grid cells or pixels, in the lateral $x$ and $y$ directions, while $M$ denotes the number of layers along the longitudinal z-axis.  To transform the continuous model into a form suitable for implementation and numerical computation, the coordinates in the x-y plane are first nondimensionalized and discretized.

Specifically, the spatial coordinates in the x-y plane are mapped to the 
interval $[-1,1]$ and discretised into $2N$ points:
\[
  x_i = -1 + \frac{2\,(i-1)}{2N-1},
  \qquad
  i = 1,2,\dots,2N.
\]
\[
  y_j = -1 + \frac{2\,(j-1)}{2N-1},
  \qquad
  j = 1,2,\dots,2N.
\]
The depth coordinate is likewise mapped to the interval $[-1,1]$ and discretised into $2M$ points:
\[
  z_k = -1 + \frac{2\,(k-1)}{2M-1},
  \qquad
  k = 1,2,\dots,2M.
\]
The resulting uniform three-dimensional grid contains $(2N)\times(2N)\times(2M)$ nodes, 
providing unambiguous indexing and convenient tensor storage. 
In the discrete domain, the ideal conical wavefront is represented by 
a Boolean tensor $\mathrm{PSF}_{i,j,k}$, a scaling factor denoted slope, 
is introduced to discretise as $\mathrm{PSF}_{i,j,k}$ follows:
\[
  \operatorname{PSF}_{i,j,k}
  = \begin{cases}
      1, & \text{if }
           \bigl(4\,\mathrm{slope}\bigr)^{2}\cdot
           \bigl(x_{i}^{2}+y_{j}^{2}\bigr) - z_{k} = 0,\\[6pt]
      0, & \text{otherwise}
    \end{cases},
\]
where $\cdot$ denotes element-wise multiplication. The prefactor $\bigl(4\,\mathrm{slope}\bigr)^{2}$ 
matches the normalised lateral distance $\sqrt{x_i^{2}+y_j^{2}} \in [-1,\,1]$ to the likewise normalised depth 
$z_k \in [-1,\,1]$ , for a wavefront propagating at $\pi /4$, the theoretical relationship is 
$z = \bigl(4\,\mathrm{slope}\bigr)\,\sqrt{x^{2}+y^{2}}$. Squaring this expression yields the 
form above, which facilitates integer-grid evaluation. To locate the wavefront correctly in 
the discrete domain, $slope$ is chosen according to the physical dimensions of the imaging region:
\[
  \text{slope}
  = \frac{w}{M \, c \,\Delta t},
\]
where $w$ is half the side length of the scanning area and $\Delta t$ is the temporal resolution.

The preceding steps map the continuous 3-D imaging volume onto a dimensionless 
discrete grid, yielding three key advantages. Scale normalisation: all 
coordinates lie in $[-1,1]$, so standard orthogonal bases and convolution 
kernels can be applied without modification. Adjustable resolution: the 
sampling density scales linearly with the parameters $N$ and $M$. Physical 
consistency: the factor slope links lateral geometric distance to the 
axial “time-depth” coordinate, ensuring that the discrete PSF aligns exactly 
with the true wavefront. Consequently, whether one employs convolution-based 
forward models, inverse solvers, or end-to-end deep networks, $\mathrm{PSF}_{i,j,k}$ 
can be used directly as the space-time propagation kernel with high computational efficiency.

To enable progressive resampling and fast convolution, we now give a systematic discrete 
formulation of the depth-direction resampling matrix $R_{z}\{ \}$, the temporal inverse-resampling 
matrix $R_{t}^{-1}\{ \}$ , and the subsequent multistage down-sampling scheme, together with their 
frequency-domain interpretation. Let the depth axis be discretised into $2M$ nodes. In several 
optical models, the spherical propagation of a plane wave permits a square-root mapping that 
folds the two-dimensional lateral index onto a one-dimensional radial depth. The matrix 
$R_{z}\{ \}$ performs this folding while maintaining amplitude consistency: a row has a non-zero 
entry only when its index is a perfect square, when it lies on a discrete radius—and 
that entry appears in column $j = \sqrt{i}$. Multiplying by $\dfrac{1}{\sqrt{i}}$ normalises the amplitude so that 
energy is conserved before and after the fold.
\[
  (R_{z})_{i,j} =
  \begin{cases}
    \dfrac{1}{\sqrt{i}}, & \text{if } j = \sqrt{i},\\[6pt]
    0,                   & \text{otherwise}.
  \end{cases}
\]
and
\[
\begin{aligned}
  R_{t}^{-1} &= R_{z}^{T}\,D_{v}^{\frac{3}{2}}, \\[6pt]
  D_{v} &= \operatorname{diag}\bigl(v_{1}, v_{2}, \dots, v_{M^{2}}\bigr).
\end{aligned}
\]
where $v_i = \left(\dfrac{c\,t_i}{2}\right)^{2}$ is the discretised temporal parameter. Excessive depth 
sampling causes the convolution dimensionality to explode. Therefore, we apply a 
binary recursive down-sampling scheme to both $R_{z}\{ \}$ and $R_{t}^{-1}\{ \}$ , ensuring that the 
spatio-temporal coupling scale coarsens in step with the overall resolution. Let
\[
  (S_k)_{i,j} =
  \begin{cases}
    \dfrac{1}{2}, & \text{if } j = 2i-1 \text{ or } j = 2i,\\[6pt]
    0,            & \text{otherwise}.
  \end{cases}
\]
the matrix averages each pair of adjacent depth (or temporal) layers into a single layer. Consequently,
\[
  R_{z}^{\text{final}}
  \;=\;
  \left(
    \prod_{k=1}^{\log_{2} M}
    \frac{1}{2}\,S_{k}
  \right)
  R_{z}^{\text{init}},
\]
\[
  R_{t}^{\text{final}}
  \;=\;
  \left(
    \prod_{k = 1}^{\log_{2} M}
    \frac{1}{2}\,S_{k}^{T}
  \right)
  R_{t}^{\text{init}}.
\]
A brute-force 3-D convolution entails $\mathcal{O}\!\bigl((2N)^{4}\,(2M)^{2}\bigr)$ operations. 
Employing the FFT converts the convolution into element-wise multiplication, 
reducing the cost to $\mathcal{O}\!\bigl((2N \times 2M)\,\log(2N \times 2M)\bigr)$. This frequency-domain formulation 
is therefore far more efficient than its spatial-domain counterpart, and the 
transient signal can be expressed as
\[
  \tau
  \;=\;
  R_{t}^{-1}\,
  \mathcal{F}_{3\mathrm{D}}^{-1} 
  \Bigl[
      \mathcal{F}_{3\mathrm{D}} \bigl(R_{z}\{p\}\bigr)
      \;\cdot\;
      \mathcal{F}_{3\mathrm{D}}(\mathrm{PSF})
  \Bigr].
\]
where $\mathcal{F}_{3D}$ denotes the three-dimensional Fourier transform and $\mathcal{F}_{3\mathrm{D}}^{-1}$ its inverse. 
By integrating the rigorously discretised operators $R_{z}\{ \}$  and $R_{t}^{-1}\{ \}$ with energy-preserving 
multilevel down-sampling and frequency-domain convolution, the transient-reconstruction 
pipeline maintains physical fidelity while fully exploiting the computational efficiency of the FFT.

Building on the foregoing model, we developed a corresponding simulation algorithm.
\begin{algorithm}
    \caption{Non-line-of-sight Simulation Pipeline}
    \label{alg:nlos_pipeline}
    
    \textbf{Input:}
    \begin{itemize}
    \item Depth map sequence: $\mathbf{D} \in \RR^{N \times N \times M}$
    \item Albedo map sequence: $\mathbf{A} \in \RR^{N \times N \times M}$ (optional)
    \item System parameters:
      \begin{itemize}
      \item Temporal resolution: $\Delta t $
      \item Spatial dimension: $N $
      \item Temporal bins: $M $
      \item PSF width: $w $
      \item Full Width Half Max of temporal jitter: $FWHM $
      \item Noise parameters: $(\sigma_{\text{blur}}=FWHM\ /2\sqrt{2ln2}, \lambda_{\text{poisson}})$
      \end{itemize}
    \end{itemize}
    
    \textbf{Output:}
    \begin{itemize}
    \item Transient measurement tensor $\TT \in \RR^{N \times N \times M\times K}$
    \end{itemize}
    
    \SetKwProg{Procedure}{Procedure}{}{}
    \Procedure{Main}{
      \textbf{System initialization:}\
      \Indp
      1.1 Compute resampling operators:\
      $\mathcal{R}_z,\mathcal{R}_t^{-1} \gets \text{ResamplingOperator}(M)$\;

      1.2 Construct PSF kernel:\
      $PSF(x,y,z) = \delta_K \left(abs\left(z - \frac{16w^2}{c^2 T_{\text{max}}^2}(x^2 + y^2)\right),0\right)$\;
      with $T_{\text{max}} = M\Delta t, \delta_K  \text{ is Kronecker delta function}$\;
      
      1.3 Precompute frequency-domain PSF:\
      $\hat{\PP} \gets \FF_{\text{3D}}(PSF)$\;
      \Indm
      
      \For{$k \gets 1$ \KwTo $K$}{
        \textbf{Frame processing:}\
        \Indp
        2.1 \textbf{Depth-to-volume mapping:}\
        $\forall (n,p) \in [1,N]^2:$\;
        $\mathbf{V}^{(k)}_{x,y,z} \gets
        \begin{cases}
        \mathbf{A}^{(k)}_{x,y}/255\cdot \delta_K (abs(z-\mathbf{D}^{(k)}_{x,y}),0) & \text{  if $\mathbf{A}$ is used} \\
        0.5\cdot \delta_K (abs(z-\mathbf{D}^{(k)}_{x,y}),0) & \text{otherwise}
        \end{cases}$\;
        
        2.2 \textbf{Transient response computation:}\
        $\mathbf{\tau}^{(k)} \gets \mathcal{R}_t^{-1} \FF_{\text{3D}}^{-1}\left[ \FF_{\text{3D}}(\mathbf{\mathcal{R}_z  V}^{(k)})\cdot \hat{\PP} \right]$\;

        2.3 \textbf{Temporal blurring:}\
        \For{$x \gets 1$ \KwTo $N$}{
          \For{$y \gets 1$ \KwTo $N$}{
            $\mathbf{\tau}^{(k)}_{x',y',:} \gets \GG_{\sigma}(\mathbf{\tau}^{(k)}_{x',y',:})$\;
            $\GG_\sigma$: Gaussian filtering with $\sigma = \sigma_{\text{blur}}$\;
          }
        }
        
        2.4 \textbf{Poisson noise injection:}\
        $\TT^{(k)} \gets \frac{1}{\sigma_{\text{n}}}\text{Poisson}\left( \mathbf{\tau}^{(k)} \cdot \sigma_{\text{n}} \right)$\;
        where $\sigma_{\text{n}}=\frac{Var(\mathbf{\tau}^{(k)})}{10^{\lambda_{\text{poisson}}/10}}$\;
        \Indm
      }
      
      \textbf{Output aggregation:}\
      $\TT \gets \bigoplus_{k=1}^K \TT^{(k)}$\;
    }
    
    
\end{algorithm}

\begin{figure}[ht!]
    \centering\includegraphics[width=12cm]{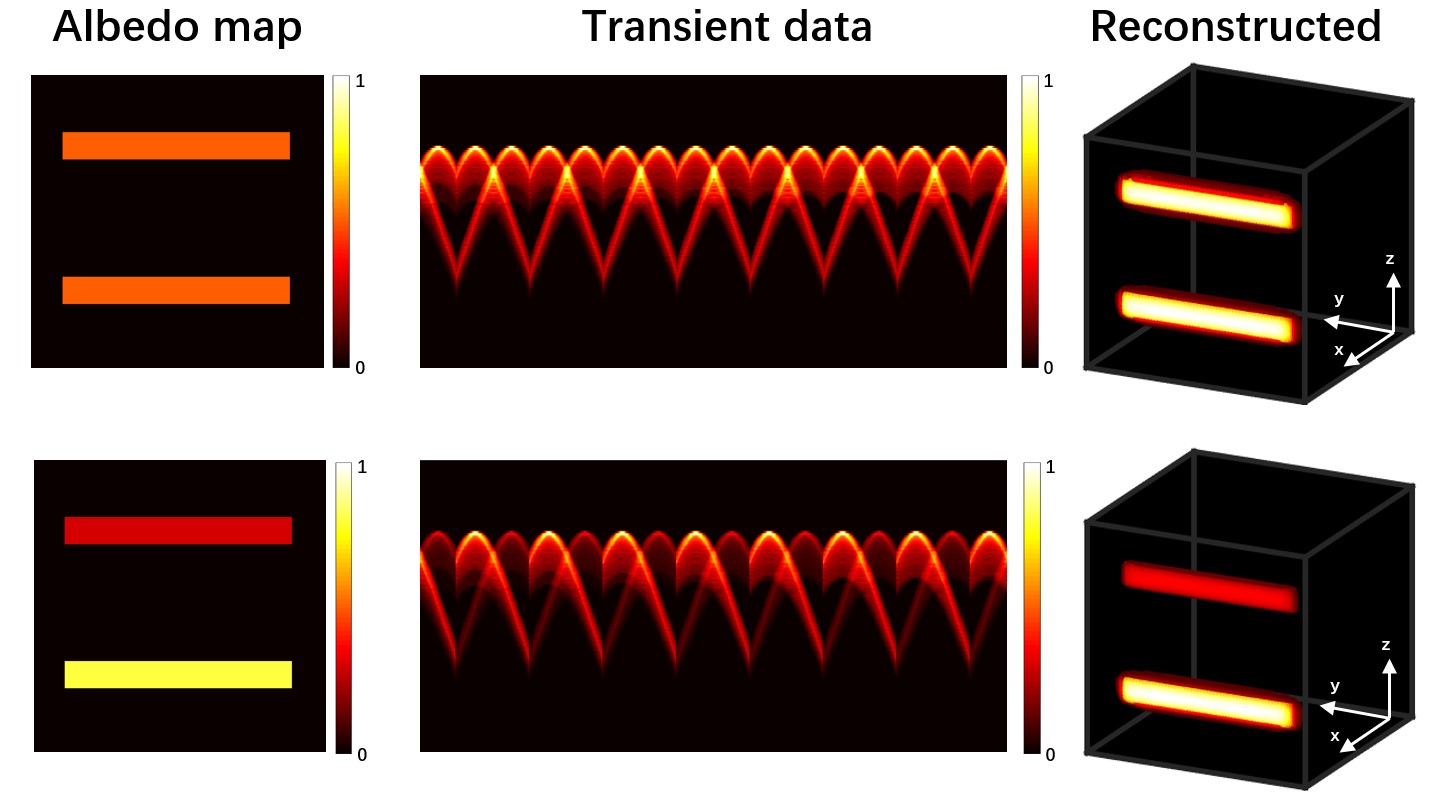}
    \caption{\textbf{Comparison of transient data and reconstruction results for targets with different albedos.}}
    \label{fig.2}
    \end{figure}

To reduce storage overhead, simplify I/O, and remain consistent with the rendering-to-reconstruction 
pipeline, we use only a depth map $D \in \mathbb{R}_{+}^{\,x\times y}$  and an albedo 
map $ A \in \mathbb{R}_{+}^{\,x\times y}$ as 
algorithm input, rather than storing a full voxel-level volume. This choice lowers the 
data dimensionality from $\mathcal{O}(xyz)$ to $\mathcal{O}(2xy)$, yielding a one- to two-order-of-magnitude 
reduction in storage and a marked decrease in disk-access latency. During computation, 
$(A,D)$ are mapped back to a discrete three-dimensional 
field $ V \in \mathbb{R}_{+}^{\,x\times y\times z}$ for the subsequent 
convolution-inversion stage, the mapping is defined as follows:
\[
  V(x,y,z)=
  \begin{cases}
    \displaystyle
    \frac{A(x,y)}{255}\,
    \delta_K\!\bigl(z - D(x,y),\,0\bigr),
    & \text{if $A$ is used},\\[6pt]
    0.5\,
    \delta_K\!\bigl(z - D(x,y),\,0\bigr),
    & \text{otherwise}.
  \end{cases}
\]
where parameter $A$ is normalised to the interval [0,1], and when $A$ is unavailable, an albedo value of 0.5 is applied. In the transient model, the voxel intensity determines the expected photon count. Voxels with low albedo (dark regions) generate fewer photons and therefore exhibit a lower SNR, whereas voxels with high albedo produce more photons and yield higher peaks in the temporal response curve. Accordingly, the assigned albedo affects both the reconstructed reflectance and the associated Poisson noise distribution. Because the detector output obeys Poisson statistics, its variance equals its mean. Voxels with low albedo thus present lower mean counts together with lower variance, whereas those with high albedo show higher means and variances. Fig. \ref{fig.2} demonstrates this behaviour by showing that simulated transient amplitudes scale with albedo, and darker regions generate weaker signals, and the same tendency is observed in the reconstructed volumes.

Detector time-jitter also influences the reconstruction and is therefore included in 
the simulation. The trigger circuitry and time-to-digital / time-to-amplitude 
converters (TDC/TAC) introduce picosecond- to nanosecond-scale uncertainty at each 
photon count, and small drifts in the laser-pulse repetition rate accumulate additional 
timing error \cite{tancock2019review,roberts2010brief,szyduczynski2023time}. In a confocal scan, every relay-wall position $(x',y')$ is recorded 
independently, with no crosstalk between adjacent pixels, so the jitter blurs only 
the temporal axis of each individual trace and does not propagate spatially. The 
timing error can be well approximated by a zero-mean Gaussian distribution. For 
the transient response at position $(x',y')$, the simulated jittered signal is:
\[
  \tau_{G}(x',y',:)
  = G_{\sigma} \bigl(\tau(x',y',:)\bigr).
\]
where $G_{\sigma}$ applies a one-dimensional Gaussian convolution with standard 
deviation $\sigma$; the kernel length k is chosen according to $\sigma$ as:
\[
  k \;=\; 2\lceil 2\sigma\rceil + 1.
\]

\begin{figure}[ht!]
    \centering\includegraphics[width=12cm]{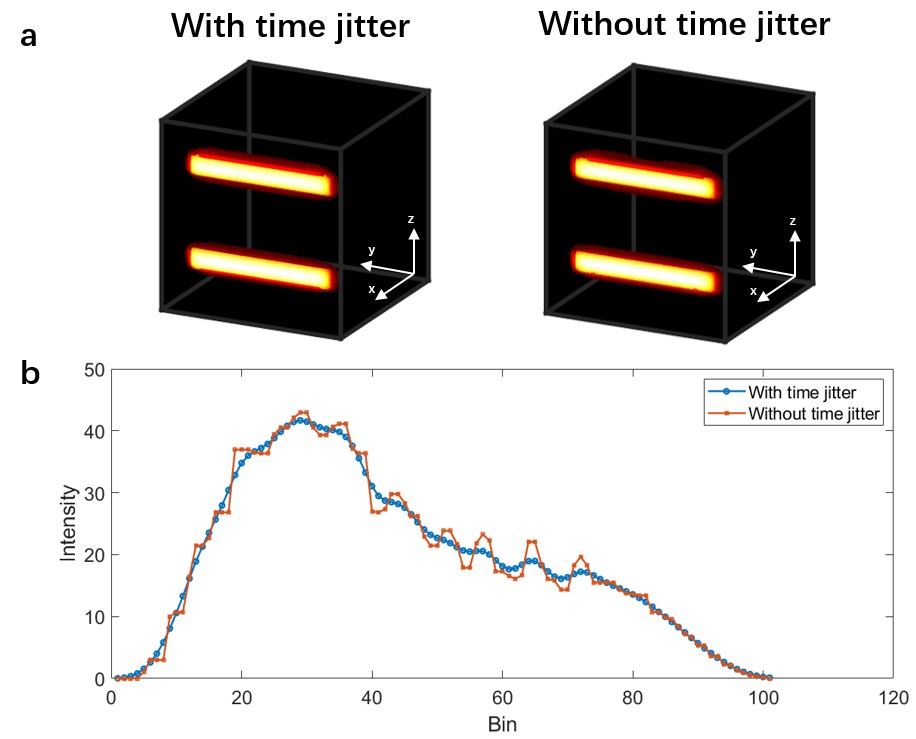}
    \caption{\textbf{Comparison of 1-D transient data and reconstruction results under different temporal jitters.}(a) Reconstruction result. (b) One‑dimensional transient histogram corresponding to a single spatial position.}
    \label{fig.3}
\end{figure}

\begin{figure}[ht!]
    \centering\includegraphics[width=12cm]{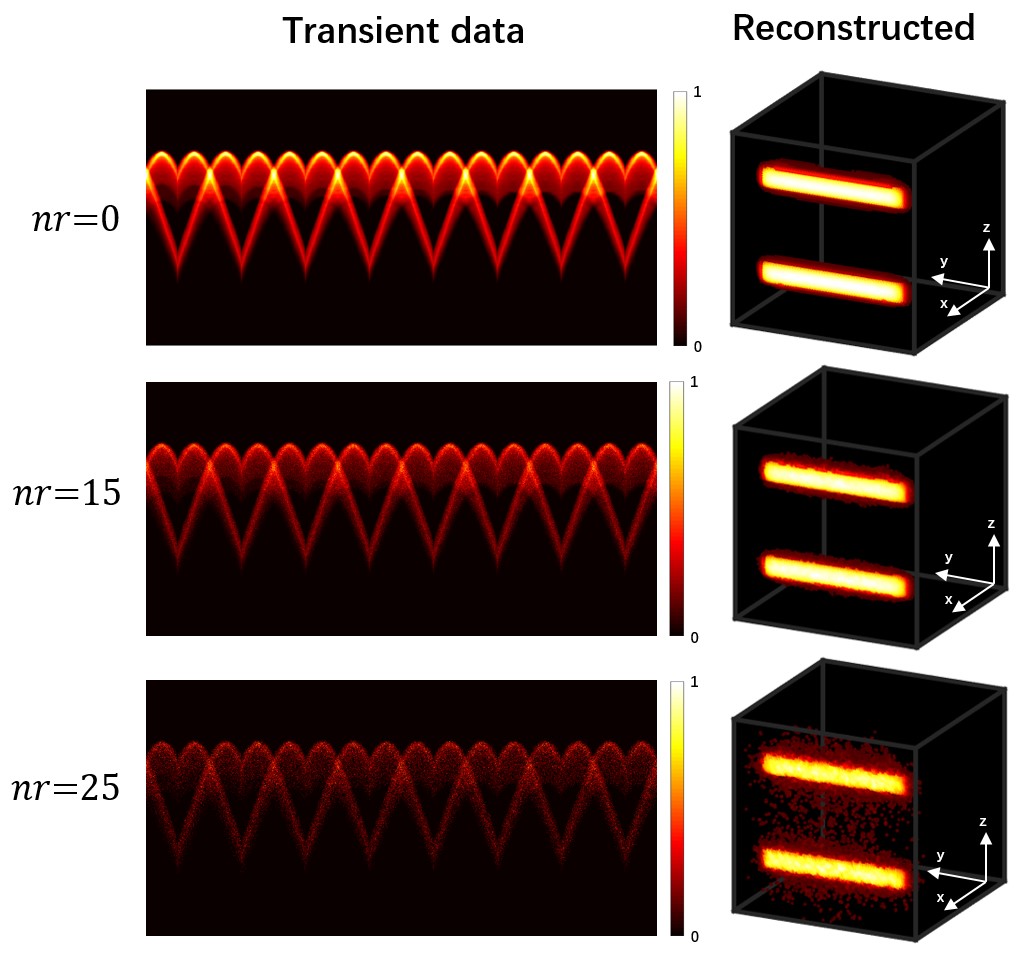}
    \caption{\textbf{Transient data and reconstruction results under different poisson noise levels.}}
    \label{fig.4}
\end{figure}

Fig. \ref{fig.3} plots the transient-intensity trace at a single relay point. Comparison of the 
curves indicates that timing jitter obscures fine temporal features, reducing the system's 
ability to reconstruct target details. 
 Single-photon counts obey Poisson 
statistics, Poisson noise is the dominant noise source in the measurement; the simulator 
explicitly models this noise as well.
\[
  T
  \;=\;
  \frac{\operatorname{Poisson} \bigl(\operatorname{Var}(\tau_{G})\,\sigma_{n}\bigr)}
       {\sigma_{n}}.
\]
the signal power is taken as the variance $\operatorname{Var}(\tau_{G})$ of the jittered trace, 
while the noise power $\sigma_{n}$ is defined as
\[
  \sigma_{n} \;=\;
  \frac{\operatorname{Var}(\tau_{G})}{10^{\text{nr}/10}}.
\]
where $nr$ is a tunable parameter that controls the noise level. Fig. \ref{fig.4} compares 
the simulated transients and the corresponding reconstructions for $nr=0$, $nr=15$, and $nr =25$. 
As $nr$ increases, Poisson noise in the transient signals becomes progressively stronger, 
ultimately leading to higher noise levels in the reconstructed volumes. Fig. \ref{fig.4} illustrates the distribution of transient data and the corresponding reconstruction results under different levels of Poisson noise. The comparison reveals that as the intensity of Poisson noise increases, the non-uniformity in the transient data becomes more pronounced, eventually leading to noticeable noise in the reconstruction.

\section{Simulation}
Recent advances in deep learning have shown considerable potential in computational imaging, especially for NLOS reconstruction. Nevertheless, most supervised networks still require large and high quality annotated data sets to reach optimal performance. Although confocal systems used by different research groups vary in scanner design, detector bandwidth, and laser wavelength, their transient measurements display a statistically similar sparse pulse distribution. This similarity indicates that a well‑trained model could generalise across hardware, provided that its training data include representative scene variations. Collecting such data remains laborious. To suppress Poisson noise and obtain high dynamic range signals suitable for reconstruction, experiments must integrate for several seconds to minutes, and each sample requires repositioning targets, adjusting occluders, and realigning optics, which makes large scale acquisition impractical. In addition, voxel level ground truth for hidden objects in natural scenes cannot be measured directly, and even laboratory targets are difficult to calibrate with sufficient accuracy. Consequently, supervised learning currently lacks reliable reference data.
\begin{figure}[ht!]
    \centering\includegraphics[width=12cm]{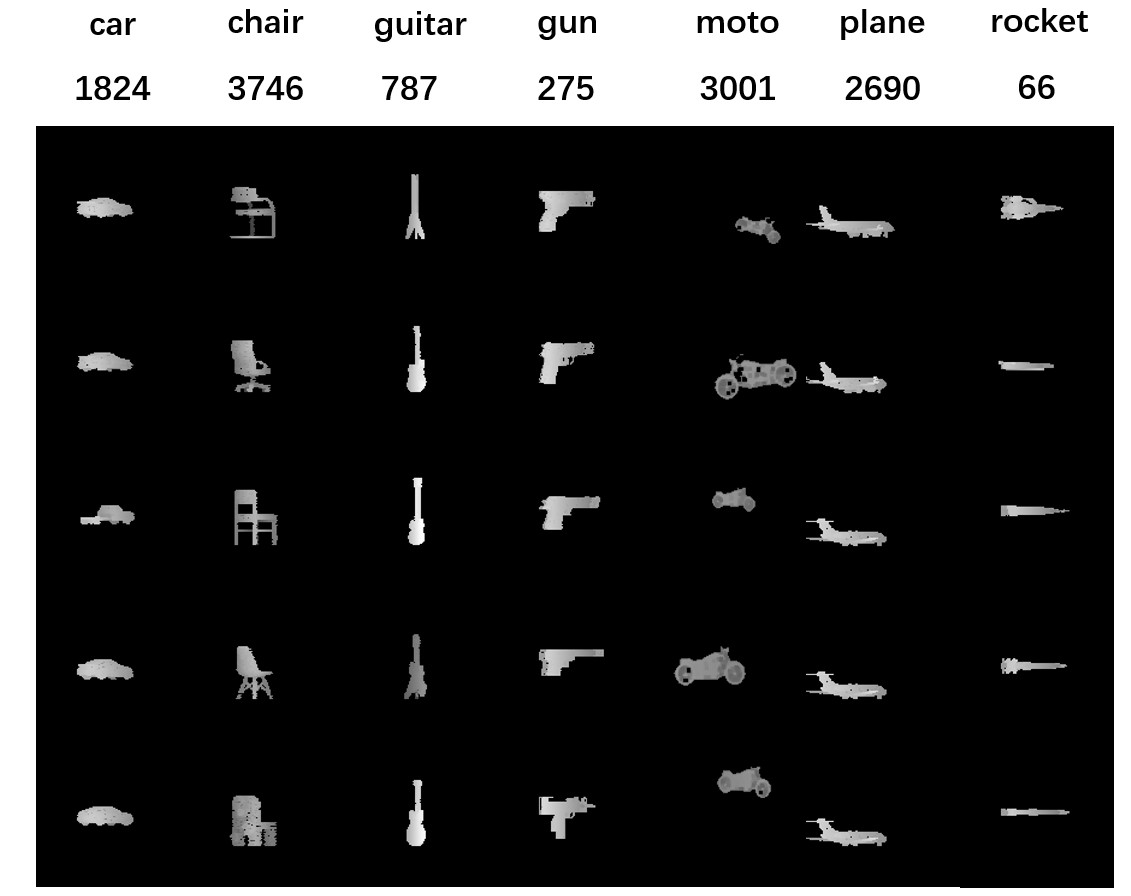}
    \caption{\textbf{Depth maps of targets from different categories.}}
    \label{fig.5}
\end{figure}

ShapeNet Core is employed as a geometric prior to sample a diverse range of object classes, and synthetic transients are generated with a rendering pipeline \cite{chang2015shapenet}. To provide a robust reference, four classical NLOS reconstruction algorithms are evaluated on the same dataset. The resulting large scale, physically consistent, and highly varied corpus not only supplies the data required for supervised learning but also offers a unified and reproducible benchmark for future studies, thereby facilitating the transition of NLOS imaging from algorithmic innovation to practical deployment.
\begin{figure}[ht!]
    \centering\includegraphics[width=12cm]{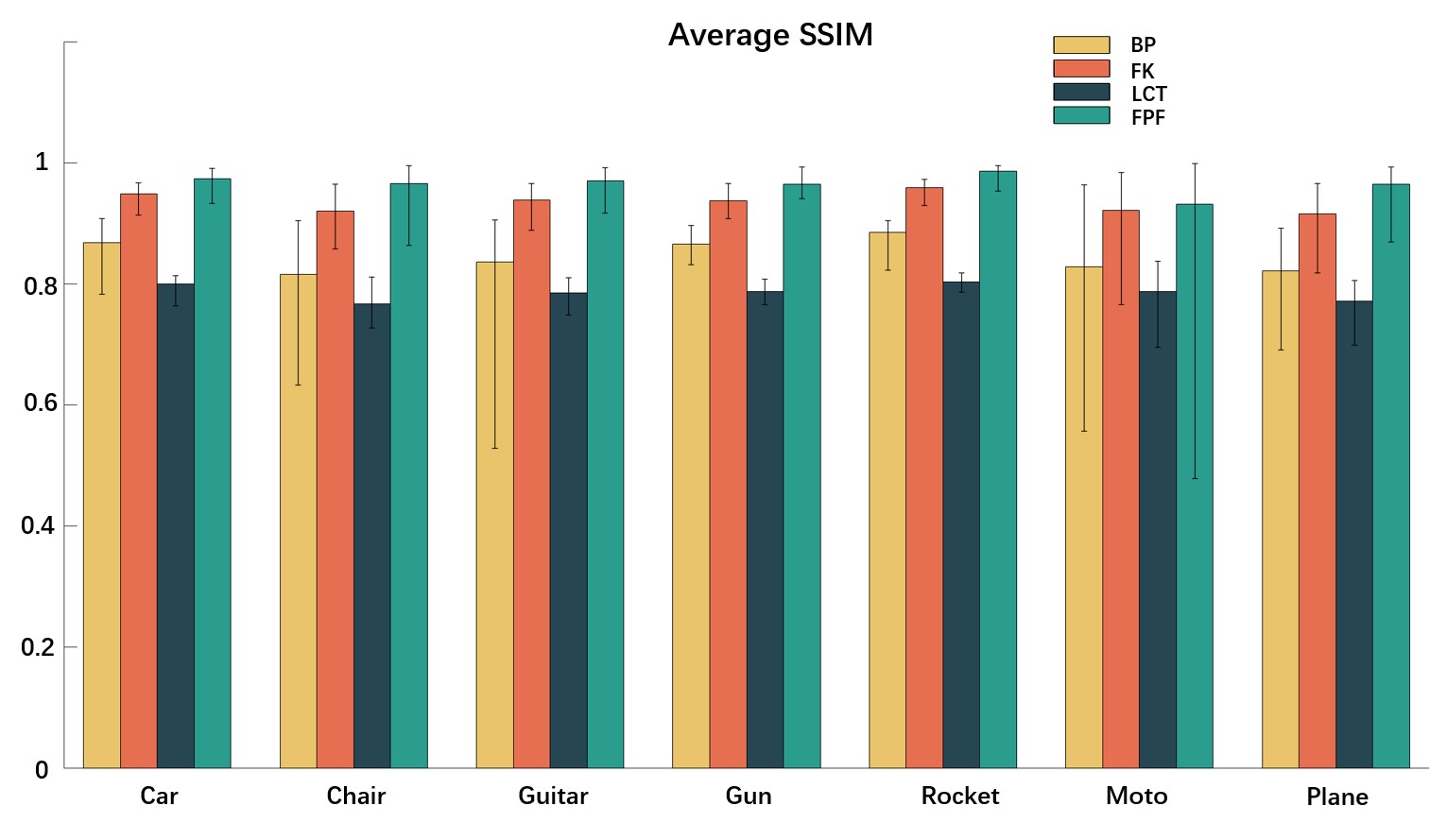}
    \caption{\textbf{SSIM comparison for targets reconstructed by different methods.}}
    \label{fig.6}
\end{figure}
\begin{figure}[ht!]
    \centering\includegraphics[width=12cm]{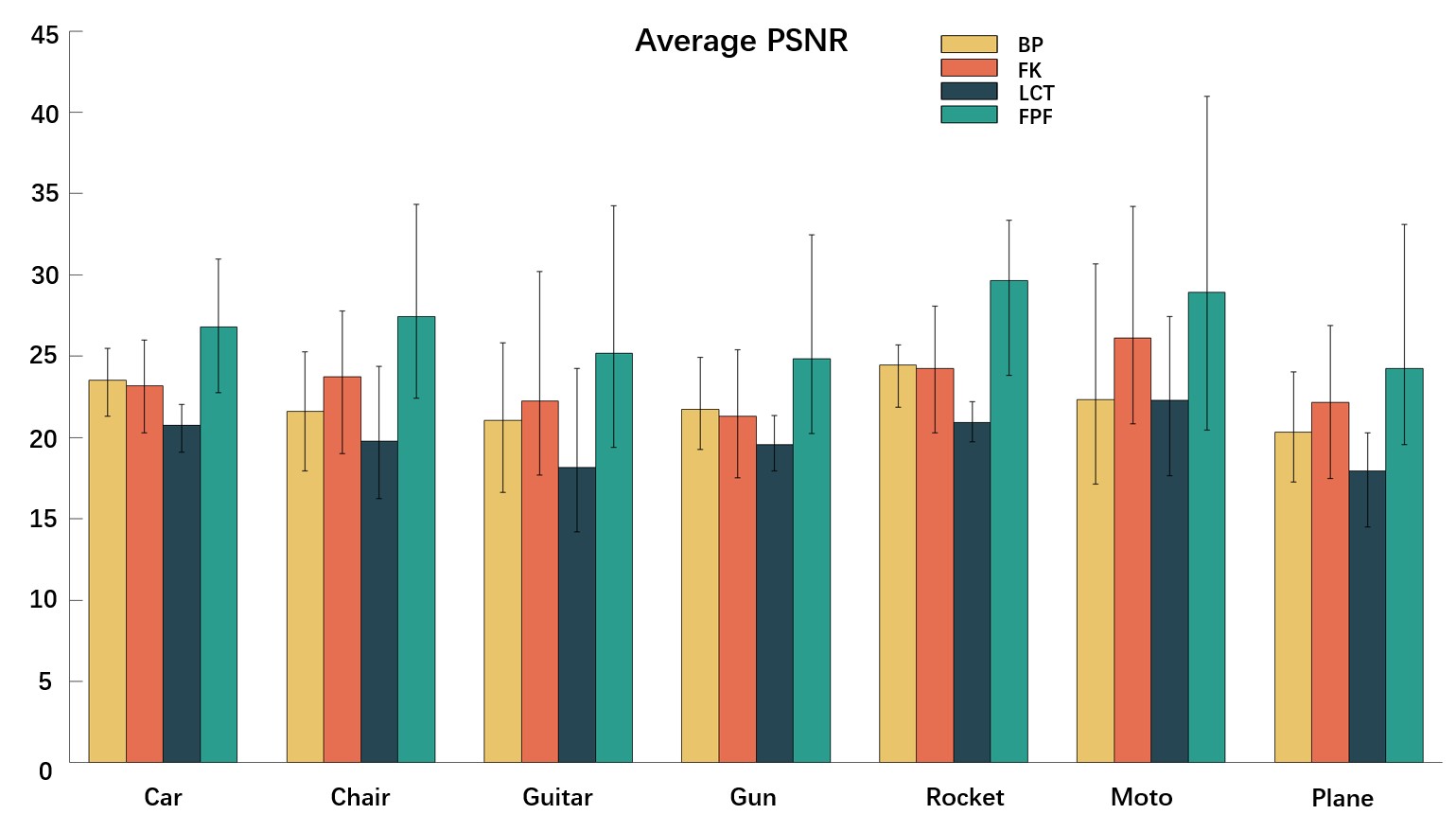}
    \caption{\textbf{PSNR comparison for targets reconstructed by different methods.}}
    \label{fig.7}
\end{figure}

To construct a training and evaluation corpus that respects NLOS confocal physics yet remains suitable for batch loading, each ShapeNet Core point cloud is orthogonally projected onto a $128\times128$ depth map, an operation that both reduces storage
and reflects the fact that a real system senses only the first visible surface. We 
select seven representative categories, such as car, chair, guitar, gun, moto, airplane, 
and rocket. And using these depth maps through the transient-rendering pipeline,  
a $128\times128\times1024$ space-time grid, a $2m\times2m$ scan area, a $32ps$ time bin, and a uniform 
albedo of 1 are used to generate transient data. Experimental noise is emulated by first applying a Poisson process 
with intensity factor $nr =2$ and then convolving each pixel's time trace with a 
$70ps$ FWHM Gaussian jitter. The resulting transient volumes serve both as benchmark 
input for classical algorithms: light-cone transform(LCT), 
phasor field(PF), f-k migration(FK), filtered backprojection(FBP) \cite{o2018confocal,liu2019non,la2020non,lindell2019wave} .
Fig. \ref{fig.6} and Fig. \ref{fig.7} show the SSIM and PSNR of reconstruction results using different algorithms, respectively.

Using the synthetic data set described above, each transient measurement was processed by all four algorithms to generate volumetric reconstructions. These reconstructions were then compared with the voxel aligned ground truth by means of two objective metrics: peak signal‑to‑noise ratio(PSNR) and structural similarity index(SSIM). PSNR measures overall numerical error, whereas SSIM correlates more closely with perceived structural fidelity; together the metrics indicate how each method balances the preservation of detail against the suppression of noise.

We also computed the peak PSNR for the reconstructions 
produced by each method.

\begin{table}[htbp]
    \centering
    \caption{Average SSIM comparison of different methods on seven target classes}
    \label{tab:ssim_comparison}
    \begin{tabular}{lccccccc}
      \toprule
      & car & chair & guitar & gun & rocket & moto & plane \\
      \midrule
      FBP  & 0.867 & 0.816 & 0.835 & 0.865 & 0.885 & 0.828 & 0.820 \\
      FK  & 0.948 & 0.920 & 0.938 & 0.937 & 0.958 & 0.921 & 0.915 \\
      LCT & 0.799 & 0.766 & 0.784 & 0.787 & 0.803 & 0.787 & 0.771 \\
      PF  & 0.973 & 0.965 & 0.970 & 0.964 & 0.985 & 0.931 & 0.963 \\
      \bottomrule
    \end{tabular}
\end{table}

\begin{table}[htbp]
    \centering
    \caption{Average PSNR comparison of different methods on seven target classes}
    \label{tab:PSNR_comparison}
    \begin{tabular}{lccccccc}
      \toprule
    & car   & chair & guitar & gun   & rocket & moto  & plane \\
      \midrule
      BP  & 23.542 & 21.617 & 21.046 & 21.742 & 24.442 & 22.324 & 20.321 \\
      FK  & 23.168 & 23.731 & 22.228 & 21.315 & 24.238 & 26.103 & 22.172 \\
      LCT & 20.773 & 19.765 & 18.147 & 19.581 & 20.946 & 22.269 & 17.939 \\
      PF  & 26.785 & 27.457 & 25.166 & 24.843 & 29.667 & 28.932 & 24.228 \\
      \bottomrule
    \end{tabular}
\end{table}

Transient data for all seven object classes, together with their corresponding depth maps, are available at \href{https://drive.google.com/drive/folders/1XW61uC3f8R3AIN-a-rtvYFBaD3JGuWWa?usp=sharing}
 {Data Repository}. A companion simulator hosted at \href{https://github.com/syjjsy/Non-Line-of-Sight-Imaging-Simulation}
 {Code Repository} enables users to upload custom depth maps or adjust simulation parameters to generate new transient data sets on demand. By rendering large scale transients for the seven ShapeNet classes and benchmarking four classical algorithms: PF, LCT, FK and FBP, andthis work provides a physically consistent, diverse training corpus and a unified, reproducible baseline.

\section{Conclusion and discussion}
An efficient framework for simulating transient data in NLOS imaging is presented. Theoretically, a rigorously discretised three‑dimensional convolution model, accelerated with the FFT and grounded in the light‑cone transform, is derived. Practically, the engine requires only depth and albedo maps, incorporates multilevel resampling, temporal jitter and Poisson noise, and synthesises data rapidly. Leveraging ShapeNet, a public data set covering seven object classes is released together with baseline results for PF, LCT, FK and FBP. The framework lowers hardware and data‑collection barriers, provides a unified and extensible platform for algorithm evaluation and deep‑learning training, and is expected to accelerate the transition of NLOS imaging from methodological innovation to real‑world deployment.
\section{Funding}
This work was supported by the National Natural Science
Foundation of China (62427803, 62031018, U23A20283);
Jiangsu Provincial Key Research and Development Program (BE2022391); the Fundamental Research Funds for
the Central Universities(30924010812);

\section{Disclosures}
The authors declare no conflicts of interest.

\section{Data availability}
Transient data for all seven object classes, together with their corresponding depth maps, are available at \href{https://drive.google.com/drive/folders/1XW61uC3f8R3AIN-a-rtvYFBaD3JGuWWa?usp=sharing}
 {Data Repository}. A companion simulator hosted at \href{https://github.com/syjjsy/Non-Line-of-Sight-Imaging-Simulation}
 {Code Repository} enables users to upload custom depth maps or adjust simulation parameters to generate new transient data sets on demand.

\bibliography{ref2}

\end{document}